\documentclass[twocolumn]{aastex63}

\usepackage{amsmath}

\newcommand{\partderiv}[2]{\frac{\partial #1}{\partial #2}}

\shorttitle{Cooling-Induced Vortex Decay}
\shortauthors{Fung \& Ono}
\graphicspath{{./}{figures/}}

\begin{document}

\title{Cooling-Induced Vortex Decay in Keplerian Disks}

\correspondingauthor{Jeffrey Fung}
\email{fung@clemson.edu}

\author{Jeffrey Fung}
\affiliation{Institute for Advanced Study, 1 Einstein Dr., Princeton, NJ 08540, USA}
\affiliation{Clemson University, 118 Kinard Laboratory,
Clemson, SC 29634, USA}

\author{Tomohiro Ono}
\affiliation{Department of Earth and Planetary Sciences, Tokyo Institute of Technology, Meguro, Tokyo 152-8551, Japan}

\begin{abstract}
Vortices are readily produced by hydrodynamical instabilities, such as the Rossby wave instability, in protoplanetary disks. However, large-scale asymmetries indicative of dust-trapping vortices are uncommon in sub-millimeter continuum observations. One possible explanation is that vortices have short lifetimes. In this paper, we explore how radiative cooling can lead to vortex decay. 
Elliptical vortices in Keplerian disks go through adiabatic heating and cooling cycles. Radiative cooling modifies these cycles and generates baroclinicity that changes the potential vorticity of the vortex. We show that the net effect is typically a spin down, or decay, of the vortex for a sub-adiabatic radial stratification.
We perform a series of two-dimensional shearing box simulations, varying the gas cooling (or relaxation) time, $t_{\rm cool}$, and initial vortex strength. We measure the vortex decay half-life, $t_{\rm half}$, and find that it can be roughly predicted by the timescale ratio $t_{\rm cool}/t_{\rm turn}$, where $t_{\rm turn}$ is the vortex turnaround time. Decay is slow in both the isothermal ($t_{\rm cool}\ll t_{\rm turn}$) and adiabatic ($t_{\rm cool}\gg t_{\rm turn}$) limits; it is fastest when $t_{\rm cool}\sim0.1\,t_{\rm turn}$, where $t_{\rm half}$ is as short as $\sim300$ orbits. At tens of au where disk rings are typically found, $t_{\rm turn}$ is likely much longer than $t_{\rm cool}$, potentially placing vortices in the fast decay regime.
\end{abstract}


\section{Introduction} \label{sec:intro}

Recent observations have uncovered annular sub-structures in numerous protoplanetary disks (e.g., \citealt{Fedele17,Fedele18,Long18,DSHARP2-rings,Long20}). Narrow rings appear to be a ubiquitous feature in these planet-forming environments. The DSHARP survey, in particular, brought to light the statistical properties of these rings (\citealt{DSHARP2-rings}). In 15 out of 18 disks, they found rings that are as narrow as $\Delta r / r \sim 0.1$, where $\Delta r$ is the ring's width, and $r$ is the ring's radius. On the other hand, only 2 out of the 18 disks showed significant azimuthal asymmetries. Large-scale asymmetries appear more common in transition disks---disks with large inner dust cavities. \cite{Francis2020} found that 9 out of 38 transition disks have signficant asymmetries. The origin of these asymmetries remain an active field of research. Some possible explanations included disk vortices \citep[e.g.,][]{vanderMarel2013,Boehler2017}, spiral arms \citep[e.g.,][]{Tang2017}, and eccentric disks \citep[][]{Ragusa2017}. Among them, vortices are a prime candidate because they can be readily formed in protoplanetary disks, so much so that it begs the question why we do not find more of them.

Narrow rings, like those found in the DSHARP survey, are vulnerable to the Rossby wave instability (RWI; \citealt{Lovelace99,Li2000}) which leads to vortex formation. In the simplest picture where we imagine the protoplanetary disk is inviscid and isothermal, RWI should operate on rings with widths comparable to or less than the disk's scale height. \cite{Ono16} showed that a pressure bump in the disk, even as small as just a few percent in amplitude, can be unstable. Moreover, because dust grains are trapped inside vortices, even a weak vortex can create a strong asymmetry in continuum emission unlikely to be missed by observations \citep{Birnstiel13}. The fact that many narrow rings appear to remain axisymmetric is surprising.

One simple resolution to this issue is to introduce some viscosity to the model. Several different groups have found that vortices decay quickly around gap edges in planet-disk simulations if the dimensionless viscosity parameter $\alpha$ (\citealt{ShakuraSunyaev73}) is larger than $10^{-3}$ (e.g., \citealt{Fung17}; \citealt{DSHARP7-planets}; \citealt{Dong18}), or if there are viscous layers present in the disk \citep{Lin2014}. While this is an easy adjustment to make in simulations, the physical interpretation of this viscosity is unclear. Gas in protoplanetary disks is not inherently viscous; hence, viscosity is typically understood as a proxy for turbulent diffusion, and the turbulence is a product of some disk instabilities. In this case, the instability in action is RWI, so the addition of any viscosity would imply that some other instabilities are operating simultaneously. It is unclear that the interaction between RWI and these instabilities, whatever they might be, is accurately modeled by the inclusion of viscosity.

Instead, we seek an alternative explanation for the prevalence of narrow rings,
and find it in radiative cooling (and heating). A gas parcel inside an elliptical vortex goes through cycles of adiabatic compression and expansion---it is adiabatically heated near its closest approach to the vortex center where gas density is highest, and cools down as it moves away. A passively heated disk should radiatively cool the gas when it is hot and heat it up when it is cold, which one can model as some form of thermal relaxation. Such a process introduces baroclinicity in the gas and changes its potential vorticity. The baroclinic term can be either positive or negative at different parts of the vortex, but it may not average out over one cycle because the gas trajectory in the vortex is elliptical and has time-varying speed. The evolution of the entropy and pressure of a gas parcel inside the vortex follows a Carnot process, which, depending on the radial stratification and the cooling time, is either converting motion into heat or entropy gradients into convection. We will later see that, in the net, this leads to the gas decelerating in angular speed and the vortex decaying. In other words, radiative cooling may be the reason why vortices decay and form narrow rings.

To better understand and quantify this decay process, we perform a series of shearing box simulations where we form vortices from RWI, and observe their thermal relaxation-induced decay. We describe our simulation setup in Section \ref{sec:method}, and in Section \ref{sec:results}, we present vortex decay rates as functions of the relaxation rate of the gas and the initial vortex strength. Finally, we discuss the implications of our findings and chart future investigations in Section \ref{sec:conclude}.

\section{Method} \label{sec:method}

We use the finite volume, Lagrangian-remap hydrodynamics code \texttt{PEnGUIn} to simulate our vortices. The two-dimensional shearing box equations in the Lagrangian frame are:
\begin{align}
\label{eq:cont_eqn}
\frac{{\rm D}\Sigma}{{\rm D}t} &= -\Sigma\left(\nabla\cdot\mathbf{v}\right) \,,\\
\label{eq:momentx_eqn}
\frac{{\rm D} v_{\rm x}}{{\rm D}t} &= -\frac{1}{\Sigma} \partderiv{P}{x} + 3x\Omega^2 + 2v_{\rm y}\Omega \,,\\
\label{eq:momenty_eqn}
\frac{{\rm D} v_{\rm y}}{{\rm D}t} &= -\frac{1}{\Sigma} \partderiv{P}{y} - 2v_{\rm x}\Omega\,,
\end{align}
where $\Sigma$ is the disk surface density, $P$ is the vertically averaged pressure, $\mathbf{v}=(v_{
\rm x}, v_{\rm y})$ is the velocity field, and $\Omega$ is the orbital frequency of the frame.
\texttt{PEnGUIn} solves the above set of equations, but with a slight modification: instead of advecting the linear y-momentum $\Sigma v_{\rm y}$, it advects the linearized angular momentum $\Sigma l\equiv \Sigma(v_{\rm y}+2x\Omega)$. The momentum equations then read:
\begin{align}
\label{eq:momentx_eqn_mod}
\frac{{\rm D} v_{\rm x}}{{\rm D}t} &= -\frac{1}{\Sigma} \partderiv{P}{x} - x\Omega^2 + 2l\Omega \,,\\
\label{eq:angmoment_eqn}
\frac{{\rm D} l}{{\rm D}t} &= -\frac{1}{\Sigma} \partderiv{P}{y} \,.
\end{align}
Thus the Coriolis term in the second equation is absorbed into the conservation of the linearized angular momentum. This treatment is analogous to tracking angular momentum instead of linear momentum in global simulations \citep{Kley98}. Our numerical tests show that this significantly improves the accuracy of our simulations.

To simulate cooling, we evolve the internal energy equation as follows:
\begin{equation}
\frac{{\rm D} e}{{\rm D}t} = -\frac{P}{\Sigma} \nabla \cdot \mathbf{v} + \frac{e_0 - e}{t_{\rm cool}}\,,\\
\label{eq:energy.eqn}
\end{equation}
where $t_{\rm cool}$ is the cooling time \footnote{This ``cooling'' term also leads to heating when $e<e_0$, so it is in fact the ``relaxation'' term. However, we choose to follow the naming convention ``cooling time'' as was done in the past, which helps underscore that, in our case, the vortices are hot, and so they are overall cooling rather than heating up. We will use the terms ``cooling'' and ``thermal relaxation'' interchangeably as long as the context is clear.}, $e=P/((\gamma-1)\Sigma)$ is the internal energy, and $e_0$ is the background specific internal energy. The second term on the right-hand side is the ``cooling'' or ``relaxation'' term that may lead to vortex decay. For convenience, we parametrize $t_{\rm cool}$ as $\beta\Omega^{-1}$, where $\beta$ is a non-dimensional number that we vary. Finally, we complete our set of equations with an adiabatic equation of state: $P = K\Sigma^{\gamma}$, where $P$ is the vertically averaged gas pressure, and the adiabatic index $\gamma$ is $1.4$.

\subsection{Boundary and Initial Conditions} \label{sec:init}

Our simulation domain covers $-20h$ to $20h$ in both $x$ and $y$ directions, where $h$ is the disk scale height. The scale height can be expressed as $c_{\rm iso}/\Omega$, where $c_{\rm iso} = \sqrt{(\gamma-1) e_0}$ is the isothermal sound speed of the background gas. In code units, $\Omega$ and $c_{\rm iso}$ both equal $1$, hence $h$ is 1 as well. The boundary is periodic in the $y$ direction, and fixed to the initial conditions in the $x$ direction. Additionally, we place wave killing zones along the $x$ boundaries between $x=\{-20h, -19h\}$ and $\{19h, 20h\}$, where all variables are gradually damped to their initial conditions following this expression:
\begin{equation}
    \frac{\partial\chi}{\partial t} = \frac{\chi(t=0)-\chi}{\Omega^{-1}} \sin^2\left(\frac{\pi|x-x_{\rm kill}|}{2h}\right)  \, .
    \label{eq:kill}
\end{equation}
$\chi$ represents all hydrodynamics variables $\Sigma$, $P$, $v_{\rm x}$, and $v_{\rm y}$. $x_{\rm kill}$ is $-19h$ or $19h$ for the inner and outer boundaries respectively.

To trigger the formation of vortices via RWI, we initialize our simulations with a Gaussian density (and pressure) peak at $x=0$. Namely, we have:
\begin{equation}
    \Sigma = \Sigma_0 \left(1 + A e^{-\frac{x^2}{2 w^2}}\right) \, ,
    \label{eq:sigma}
\end{equation}
where $\Sigma_0=1$ is a normalization for the surface density, $A$ is a parameter we vary to create vortices of different strengths, and $w=0.5h$ is the width of the peak. Gas pressure is initialized as $P = c_{\rm iso}^2\Sigma$.

The velocity field assumes hydrostatic equilibrium plus a small initial perturbation to set the stage for the growth of RWI modes:
\begin{align}
\label{eq:vx}
v_{\rm x} &= \frac{\epsilon}{2}\left(\sin{\frac{2\pi y}{5h}}-\sin{\frac{4\pi y}{5h}}\right) e^{-\frac{1}{2}\left(\frac{x}{h}\right)^2} \,,\\
\label{eq:vy}
v_{\rm y} &= -\frac{3\Omega x}{2} + \frac{1}{2\Omega\Sigma}\frac{{\rm d} P}{{\rm d}x} \,.
\end{align}
where $\epsilon=10^{-5}$ is the amplitude of the initial perturbation.

We choose values of $A$ based on a linear analysis of the stability of our initial profiles. We sample four values for $A$: $0.054$, $0.066$, $0.098$, and $0.16$; our analysis suggests that they are all unstable, where the fastest growing unstable modes have modes numbers 1, 1, 2, and 2, with corresponding growth rates of $0.012\Omega$, $0.025\Omega$, $0.06\Omega$, and $0.12\Omega$. Initially, the number of vortices formed equals the mode number, but they quickly merge, leaving only one vortex in our simulation domain.

The other parameter we vary is $\beta$. We sample seven different values: $0.1$, $0.3$, $1$, $3$, $10$, $30$, and $100$. Together with our four values for $A$, this gives a total of twenty-eight simulations, summarized in Table \ref{tab:all}.

\begin{table*}
\begin{center}
\caption{List of simulations.}
\label{tab:all}
\begin{tabular}{llllll}
\tableline \tableline
$A$ & $\beta$ & $t_{0}$ & $t_{\rm turn}(t=t_0)$ & $t_{\rm half}$ \\
~ & ~ & [orbits] & [$\Omega^{-1}$] & [$\Omega^{-1}$] \\
\hline 
\hline
0.054 & 0.1 & 550 & $2.2\times10^3$ & -- \\
0.054 & 0.3 & 530 & $2.3\times10^3$ & --\\
0.054 & 1 & 520 & $2.3\times10^3$ & --\\
0.054 & 3 & 520 & $2.0\times10^2$ & $9.6\times10^3$ \\
0.054 & 10 & 520 & $1.9\times10^2$ & $2.8\times10^3$\\
0.054 & 30 & 490 & $1.8\times10^2$ & $2.4\times10^3$\\
0.054 & 100 & 520 & $1.8\times10^2$ & $3.9\times10^3$\\
\hline 
\hline
0.066 & 0.1 & 330 & $1.2\times10^2$ & -- \\
0.066 & 0.3 & 330 & $1.3\times10^2$ & --\\
0.066 & 1 & 300 & $1.3\times10^2$ & $8.9\times10^3$\\
0.066 & 3 & 270 & $1.2\times10^2$ & $3.0\times10^3$ \\
0.066 & 10 & 250 & $1.2\times10^2$ & $1.8\times10^3$\\
0.066 & 30 & 300 & $1.2\times10^2$ & $2.8\times10^3$\\
0.066 & 100 & 300 & $1.1\times10^2$ & $5.0\times10^3$\\
\hline 
\hline
0.098 & 0.1 & 180 & $6.2\times10^2$ & $4.6\times10^4$\\
0.098 & 0.3 & 180 & $5.9\times10^1$ & $1.0\times10^4$\\
0.098 & 1 & 180 & $6.0\times10^1$ & $3.5\times10^3$\\
0.098 & 3 & 180 & $6.0\times10^1$ & $1.8\times10^3$ \\
0.098 & 10 & 180 & $6.6\times10^1$ & $2.4\times10^3$\\
0.098 & 30 & 180 & $7.0\times10^1$ & $3.7\times10^3$\\
0.098 & 100 & 180 & $6.3\times10^1$ & $9.4\times10^3$\\
\hline 
\hline
0.16 & 0.1 & 100 & $3.1\times10^1$ & -- \\
0.16 & 0.3 & 100 & $3.4\times10^1$ & --\\
0.16 & 1 & 100 & $3.3\times10^1$ & $5.4\times10^3$ \\
0.16 & 3 & 100 & $3.2\times10^1$ & $2.2\times10^3$\\
0.16 & 10 & 100 & $3.4\times10^1$ & $2.3\times10^3$\\
0.16 & 30 & 100 & $3.2\times10^1$ & $1.2\times10^4$\\
0.16 & 100 & 100 & $3.5\times10^1$ & $3.5\times10^4$\\
\tableline \tableline
\end{tabular}
\end{center}
\end{table*}

\subsection{Metrics} \label{sec:metrics}

There are a number of properties we aim to extract from our simulated vortices. Here we describe our metrics.

Operating in a coordinate system where the vortex center is located at the origin, the vortex core has a profile such that the magnitude of $v_{\rm x}$ along the $y$-axis is proportional to the distance to the vortex center, and falls back to zero outside the core. Where $|v_{\rm x}|$ reaches its maximum is therefore roughly the edge of the core. We define the semi-major axis of the vortex core, denoted as ``$a$'', to be half the distance between the two maxima in $|v_{\rm x}|$ along the $y$-axis. The characteristic rotation speed  $v_{\rm rot}$ of the vortex is then equal to $|v_{\rm x}|$ averaged between $\{x,\,y\}=\{0,\,a\}$ and $\{x,\,y\}=\{0,\,-a\}$.

There are three key timescales in the systems we simulate. One is the dynamical time, $\Omega^{-1}$; another is the cooling time, $\beta\Omega^{-1}$; and the third is the vortex turnaround time, $t_{\rm turn}$. 
To measure $t_{\rm turn}$, we first measure the semi-minor axis $b$ of our vortices (not just the core as in our measurements of $a$). For a given snapshot, we integrate the velocity field to obtain streamlines, and locate the streamline that revolves around the vortex center at the largest separation. Where that streamline intersects the $x$-axis gives us $b$. Then, $t_{\rm turn}$ is the time it takes for the streamline crossing $\{x,\,y\}=\{b/2,\,0\}$ to complete one turn. We choose $b/2$ to ensure the streamline is sufficiently close to the vortex core---$t_{\rm turn}$ becomes very long as it approaches $\{x,\,y\}=\{b,\,0\}$, which is formally at the separatrix of a steady-state vortex where the time for one revolution becomes infinite.

Finally, we devise a metric to quantify vortex strength. First, consider the geostrophic balance in vortex cores:
\begin{equation}
    \frac{1}{\Sigma}\frac{{\rm d} P}{{\rm d} r} = \frac{v_{\phi} \Omega}{2}   + \frac{v_{\phi}^2}{r}\, ,
    \label{eq:balance}
\end{equation}
where $r$ is the distance to the vortex center and $v_{\phi}$ is the azimuthal speed around the vortex center. We can evaluate this equation approximately using the characteristic size $a$ and speed $v_{\rm rot}$ of the vortex: 
\begin{equation}
    \frac{\Delta P}{\Sigma} = \frac{a v_{\rm rot}\Omega}{2} - v_{\rm rot}^2\, .
    \label{eq:balance_appr}
\end{equation}
We have approximated ${\rm d}P/{\rm d}r$ as $-\Delta P / a$, where $\Delta P$ is the pressure difference between the vortex center and the background gas, and substituted $r$ with $a$ and $v_{\phi}$ with $-v_{\rm rot}$. We now define a dimensionless variable $\eta$:
\begin{equation}
    \eta \equiv \frac{1}{(\gamma-1)e_0}\left(\frac{a v_{\rm rot}\Omega}{2}  - v_{\rm rot}^2\right)\, ,
    \label{eq:eta}
\end{equation}
which quantifies pressure support in the vortex. For the kind of vortices we study (i.e. those in geostrophic balance), it is a good measurement of vortex strength and captures the effects of cooling well. It should not be generalized to represent strength for all kinds of vortices, however. For instance, a vortex in inertial balance, where Coriolis and centrifugal forces balance each other, would have no pressure support and $\eta$ equaling zero.

\subsection{Resolution} \label{sec:res}

\begin{figure}
    \centering
    \includegraphics[width=0.47\textwidth]{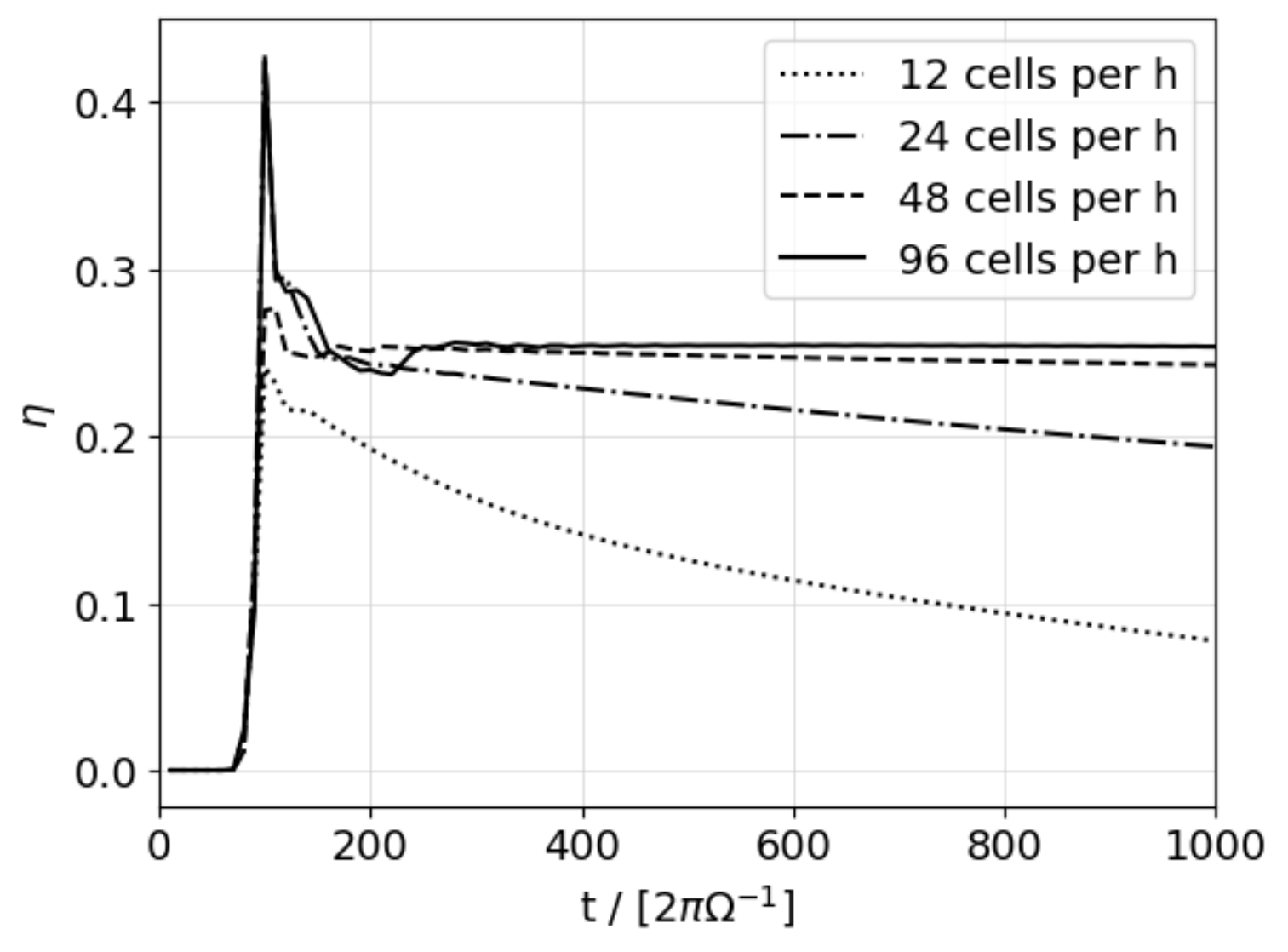}
    \caption{$\eta$ as a function of time at 4 different resolutions where $A=0.098$. The simulations are all adiabatic without cooling, and hence the vortex does not decay when resolution is sufficiently high. At our fiducial resolution, 48 cells per $h$, numerical diffusion weakens the vortex only by about $3\%$ over 800 orbits.}
    \label{fig:res}
\end{figure}

Since our goal is to measure vortex decay, it is essential that our simulations have sufficient resolution such that decay caused by numerical diffusion is insignificant. Figure \ref{fig:res} plots $\eta$ evolution at four different resolutions in the case when there is no cooling ($\beta=\infty$). We do not expect the vortex to decay in this case, and indeed we find that $\eta$ converges to a constant in time as resolution increases. At 48 cells per $h$, $\eta$ decreases by only about $3\%$ over 800 orbits. We therefore choose 48 cells per $h$ as the resolution for our runs.

\section{Results} \label{sec:results}

\begin{figure*}
    \centering
    \includegraphics[width=0.9\textwidth]{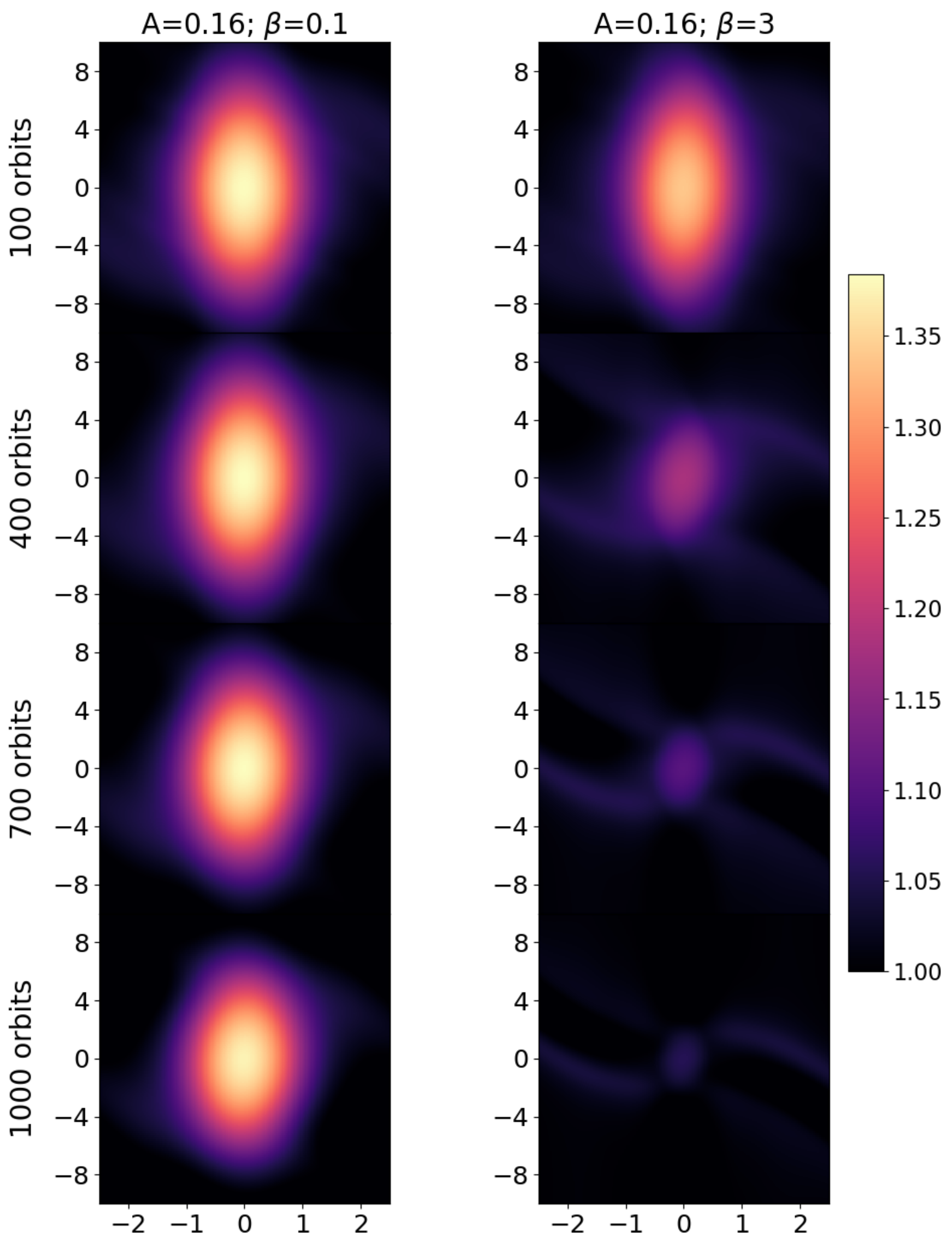}
    \caption{Surface density snapshots for two models, $\{A,\beta\}=\{0.16,0.1\}$ and $\{0.16,3\}$, at four different times. When the cooling time is short, $\beta=0.1$ on the left, the vortex is nearly isothermal to start with and subsequently experiences little decay. But with a longer cooling time, $\beta=3$, we find the vortex rapidly decays on the right.}
    \label{fig:2Dden}
\end{figure*}

\begin{figure*}
    \centering
    \includegraphics[width=0.9\textwidth]{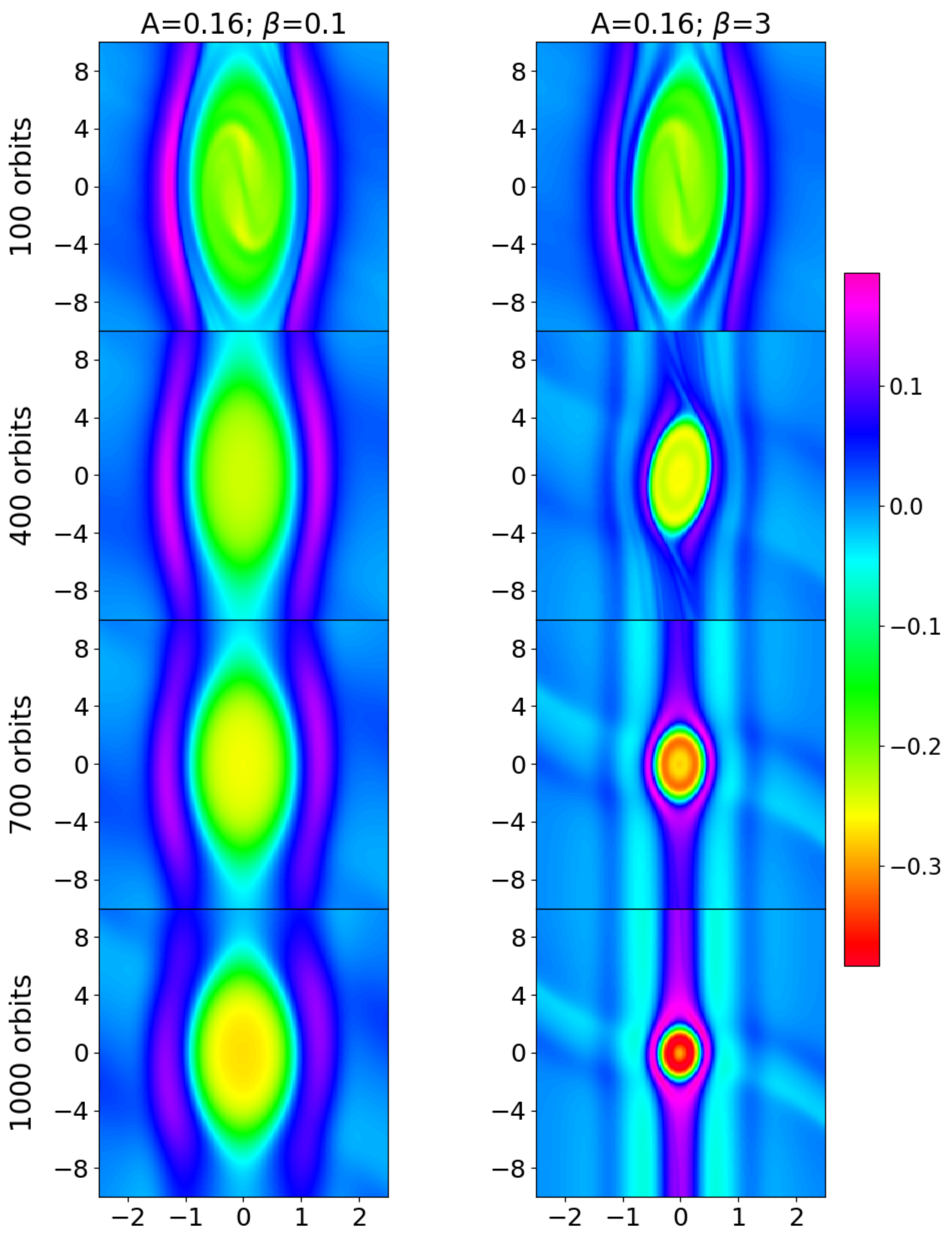}
    \caption{Same as Figure \ref{fig:2Dden} but showing vorticity snapshots rather than surface density. Color plots $\omega-\omega_{\rm K}$, where $\omega$ is the vorticity and $\omega_{\rm K}=-1.5\,\Omega$ is the vorticity of the background Keplerian shear. Vorticity in the vortex core become more negative over time, indicating a faster anticyclonic spin. This is a consequence of the core losing mass (see Section \ref{sec:results}).}
    \label{fig:2Dvor}
\end{figure*}

\begin{figure*}
    \centering
    \includegraphics[width=0.99\textwidth]{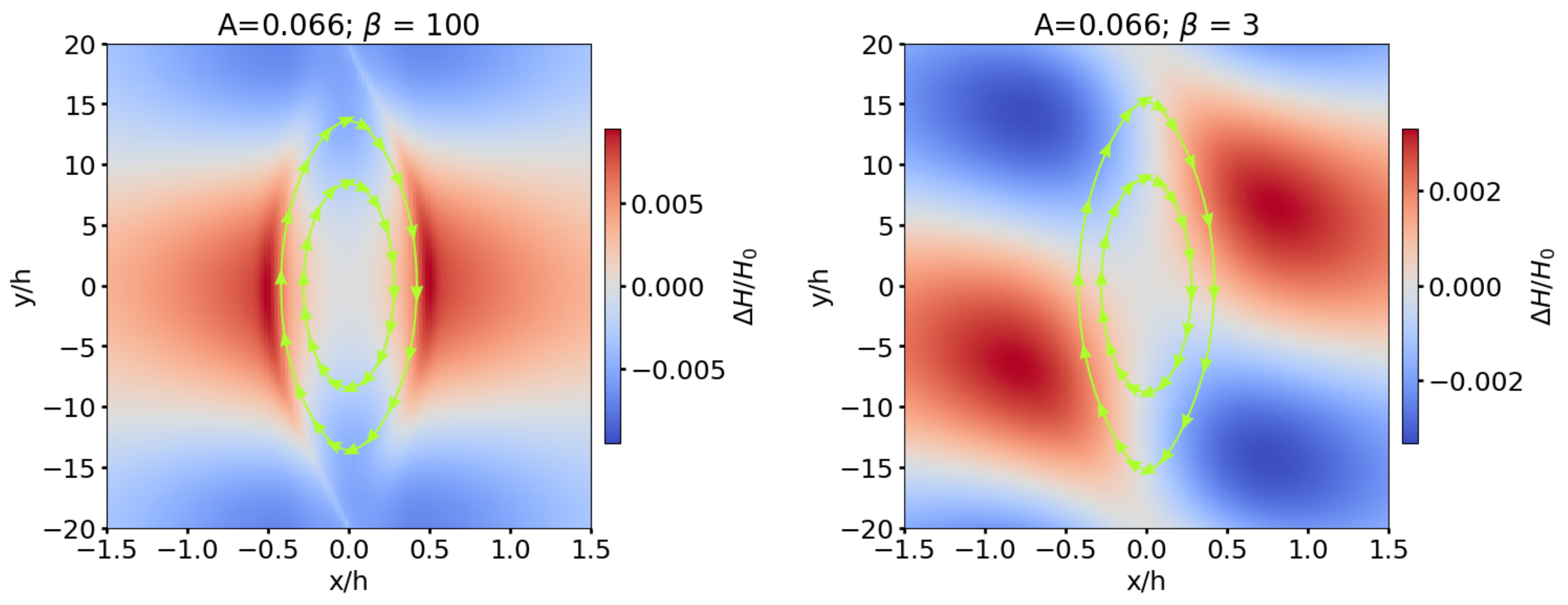}
    \caption{Enthalpy distribution in two models, $\{A,\beta\}=\{0.066,100\}$ and $\{0.066,3\}$, taken at $t=300$ orbits. Two vortex streamlines are over-plotted. The arrows are separated by a constant interval of time $\sim 7 \Omega^{-1}$, which shows the gas speed is slower further away from the vortex center. When cooling is slow (left), the enthalpy profile is left-right symmetric, following the adiabatic compression and expansion of the gas. With faster cooling (right), the profile is significantly altered, showing strong gradients near the $x$- and $y$-axis. These profiles are qualitatively representative of our models with either slow cooling (left) or moderate cooling (right). Fast cooling ($\beta<1$) resembles the right panel, but with much weaker amplitudes.}
    \label{fig:enthalpy}
\end{figure*}

\begin{figure*}
    \centering
    \includegraphics[width=0.99\textwidth]{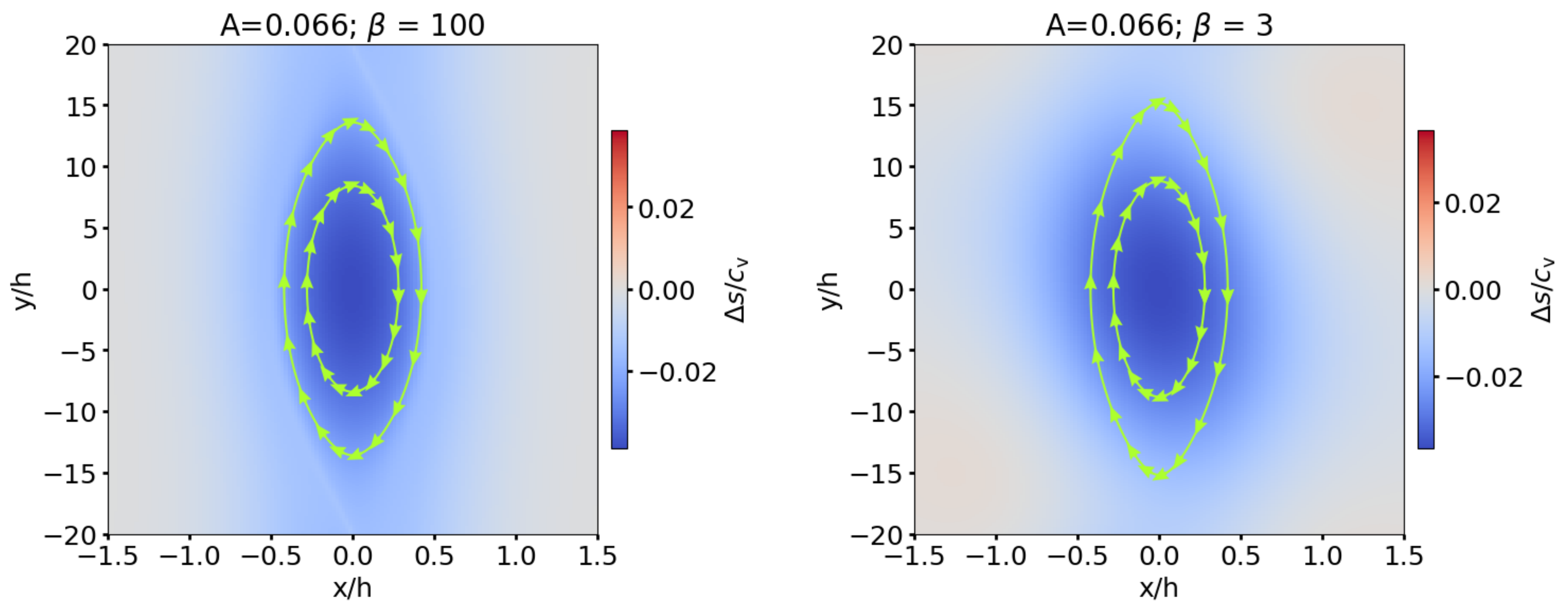}
    \caption{Same as Figure \ref{fig:enthalpy}, but plotting the entropy change $\Delta s$. As shown in these two examples, vortices in all cases have a lower entropy than their surroundings. This is because anticyclonic vortices are high pressure systems being cooled by thermal relaxation. The outskirts of the vortices have slightly varying entropy profiles depending on their $\beta$ values, but generally, the entropy gradient is mostly radial.}
    \label{fig:entropy}
\end{figure*}

\begin{figure*}
    \centering
    \includegraphics[width=0.99\textwidth]{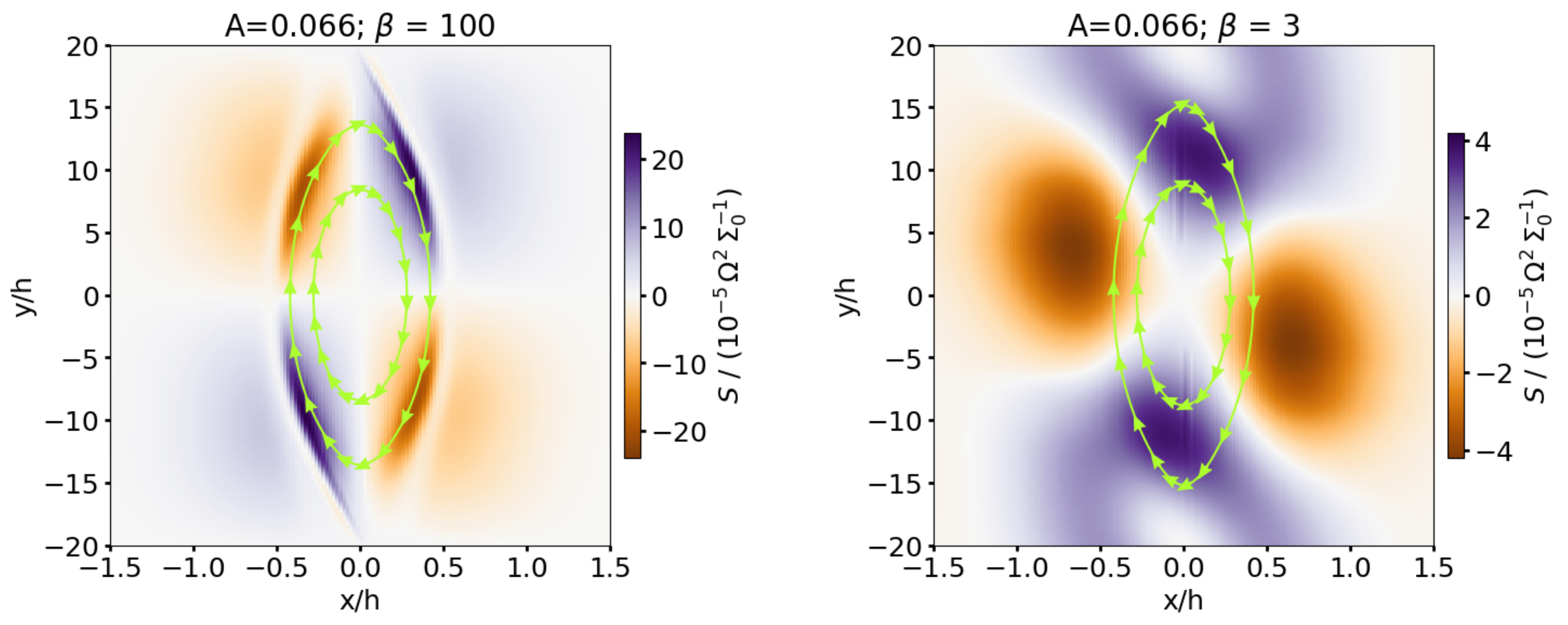}
    \caption{Same as Figure \ref{fig:enthalpy}, but plotting the profile of the potential vorticity source term $S$ (Equations \ref{eq:baroclinic_1} and \ref{eq:baroclinic_2}). When cooling is slower(left), the magnitude of $S$ is larger since there are larger variations in enthalpy (Figure \ref{fig:enthalpy}), but the positive and negative values of $S$ largely cancel over one turn around the vortex. With faster cooling (right), that symmetry is broken and produces a net positive $S$ in one turn.}
    \label{fig:baroclinic}
\end{figure*}

Figures \ref{fig:2Dden} and \ref{fig:2Dvor} plot the density and vorticity, respectively, from two representative simulations, illustrating one example of no decay on the left and one example of fast decay on the right. Clearly, the decay rate is sensitive to the cooling time $\beta\Omega^{-1}$.
To better understand vortex decay, we analyze the dynamics illustrated in Figures \ref{fig:enthalpy}, \ref{fig:entropy} and \ref{fig:baroclinic}. Figure \ref{fig:enthalpy} plots the relative enthalpy change $(H-H_0)/H_0$, where the enthalpy $H$ is $(\gamma/(\gamma-1))P/\Sigma$, and the background enthalpy $H_0$ is $(\gamma/(\gamma-1))c_{\rm iso}^2$. Figure \ref{fig:entropy} plots the entropy change $(s-s_0)/c_{\rm v}$, where $s-s_0=c_{\rm v}\ln((P/c_{\rm iso}^2 \Sigma_0)(\Sigma/\Sigma_0)^{-\gamma})$, and $c_{\rm v}$ is the specific heat at constant volume. Figure \ref{fig:baroclinic} plots the potential vorticity source term due to baroclinicity:
\begin{equation}
    S = \frac{\nabla \Sigma \times \nabla P}{\Sigma^3} \cdot \hat{z}\, ,
    \label{eq:baroclinic_1}
\end{equation}
which can also be written as:
\begin{equation}
    S = \frac{1}{c_{\rm p}}\frac{\nabla H \times \nabla s}{\Sigma} \cdot \hat{z}\, ,
    \label{eq:baroclinic_2}
\end{equation}
where $c_{\rm p}$ is the specific heat at constant pressure \footnote{The absolute values of the specific heats are irrelevant here, as long as their ratio is fixed by $\gamma=c_{\rm p}/c_{\rm v}$.}. This source term is technically a vector quantity, but since our models are 2D, its only non-zero component is along the $\hat{z}$ direction. In all three figures, we over-plot vortex streamlines.

$S$ changes the potential vorticity $PV = (\nabla\times\mathbf{v}+2\Omega)/\Sigma$, which is otherwise conserved along streamlines in our models. We also denote $\omega=\nabla\times\mathbf{v}$, the vorticity of the gas in the frame of our shearing box. Since vortices in a Keplerian shear are always anticyclonic, $\omega$ is always negative; the total vorticity $\omega+2\Omega$, on the other hand, is positive. This is because $|\omega|$ scales with $v_{\rm rot}/h$, and the vortex rotation speed $v_{\rm rot}$ is subsonic (supersonic vortices would dissipate quickly from shocks), hence $v_{\rm rot}/h<\Omega$.

When $\beta=100$, shown on the left panels of Figures \ref{fig:enthalpy}, \ref{fig:entropy}, and \ref{fig:baroclinic}, enthalpy variations are mostly due to adiabatic compression and expansion. Thermal relaxation plays a negligible role in comparison. The overall cooling of the vortex creates a radial entropy gradient, which, combined with the azimuthal variations in the enthalpy, produces a finite baroclinic term $S$. Despite that, because the adiabatic heating and cooling patterns are symmetric along vortex streamlines, positive and negative $S$ largely cancel and the net change is small.

When $\beta=3$, shown on the right panels of Figures \ref{fig:enthalpy}, \ref{fig:entropy}, and \ref{fig:baroclinic}, enthalpy variations are strongly modified by thermal relaxation. Heating by relaxation is strongest where the gas is adiabatically cooled the most, which occurs near $x=0$. Similarly, cooling by relaxation is strongest near $y=0$. As a result, the symmetry between positive and negative $S$ is now broken because a parcel of gas in a vortex has different speeds near the vortex's major axis $(y=0)$ and its minor axis $(x=0)$. To illustrate gas speeds, the arrows along the streamlines in these figures are separated by a constant interval of time, showing that the gas is moving slower when it is further away from the vortex center. The gas spends significantly longer where enthalpy is increasing along its path, which is also where $S$ is positive. As a result, it spins down, weakening the Coriolis force that holds the vortex together and leading to an outflow of mass, as seen in Figure \ref{fig:2Dden}.

Interestingly, not the entire vortex spins down. In Figure \ref{fig:2Dvor}, we see that the vortex core in fact spins up. This is because the vortex center is less elliptical and experiences less variation in enthalpy. As gas in the outskirts of the vortex flows away, the vortex core loses pressure support around it, which induces a radial outflow inside the core as well. In the absence of any significant source term $S$, such radial motion is torqued by the Coriolis force into an anticyclonic spin. Alternatively, it is also straightforward to see that when $PV$ is conserved, a decrease in $\Sigma$ must lead to a reduction in the total vorticity $\omega+2\Omega$, and a more negative $\omega$ is a faster anticyclonic spin. Despite the spin up, the vortex core is losing mass and so decay occurs in the core as well.

\begin{figure*}
    \centering
    \includegraphics[width=0.99\textwidth]{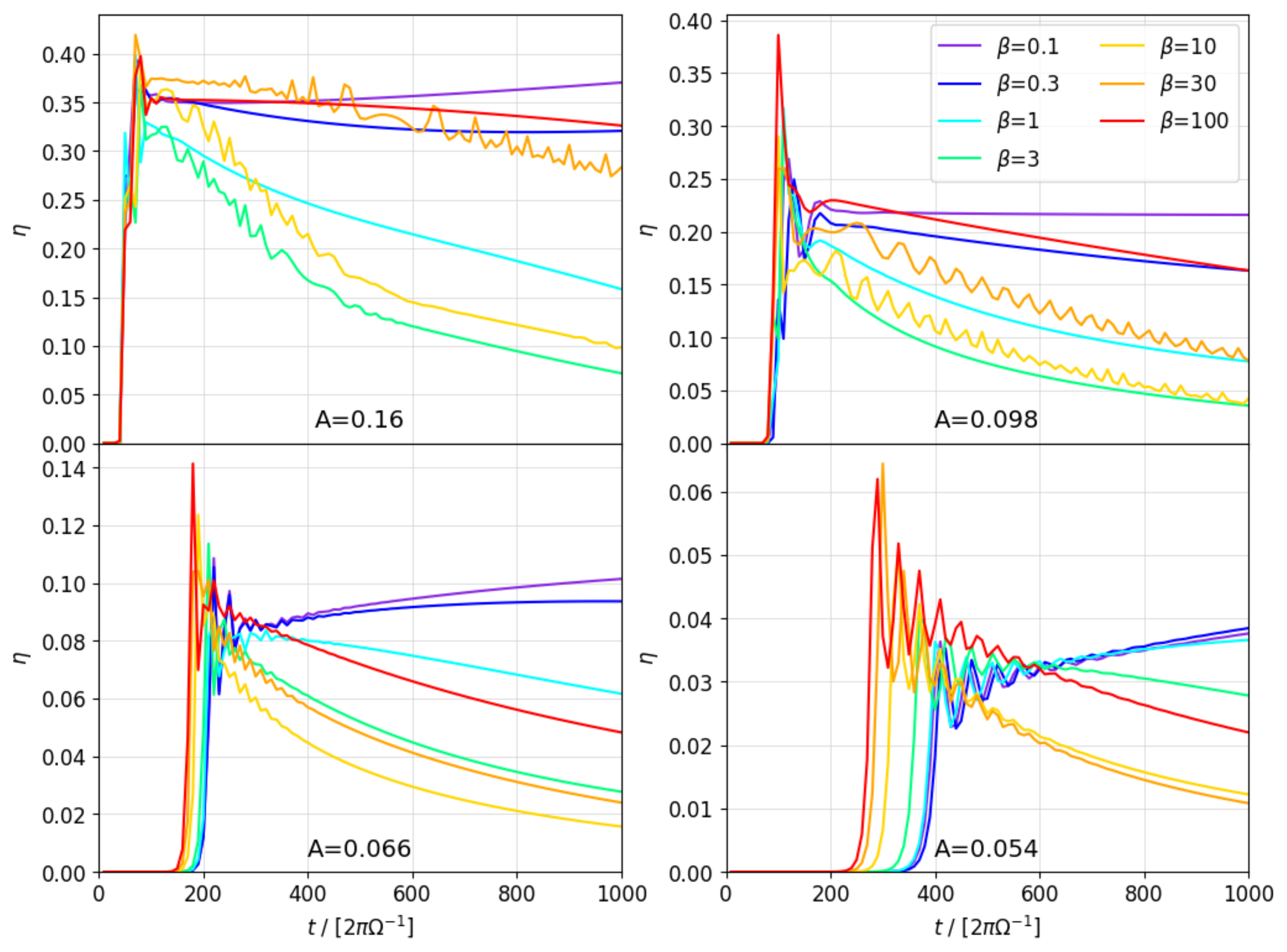}
    \caption{$\eta$ as a function of time for all 28 simulations, separated into 4 panels by the amplitude of the initial Gaussian peak $A$. Larger $A$ produces stronger vortices (larger $\eta$) that also form more quickly. For our smallest value of $A$, it takes 200 to 400 orbits for RWI to reach non-linearity and form a vortex. Faster cooling also appears to lengthen the process. After the vortices are formed, cooling generally leads to their decay, with some exceptions when $\beta\lesssim 1$. The decay is roughly exponential, allowing us to measure the half-life of those vortices.}
    \label{fig:eta}
\end{figure*}

Figure \ref{fig:eta} plots the evolution of $\eta$ in all of our simulations. As expected, $\eta$ generally decreases over time. Its decay rate is roughly exponential---although, our simulations do not cover a wide enough time range to confirm the exponential nature of the decay---and is fastest when $\beta$ is neither very large nor very small.

We have seen that for cooling to play a role, it needs to substantially alter the adiabatic compression and expansion cycles around elliptical vortices. Clearly, cooling cannot be too slow ($\beta$ too large) for this to occur. On the other hand, 
cooling cannot be too fast ($\beta$ too small) either, otherwise variations in $H$ would be heavily suppressed, which would diminish the $PV$ source term $S$ and slow down decay.

\begin{figure*}
    \centering
    \includegraphics[width=0.99\textwidth]{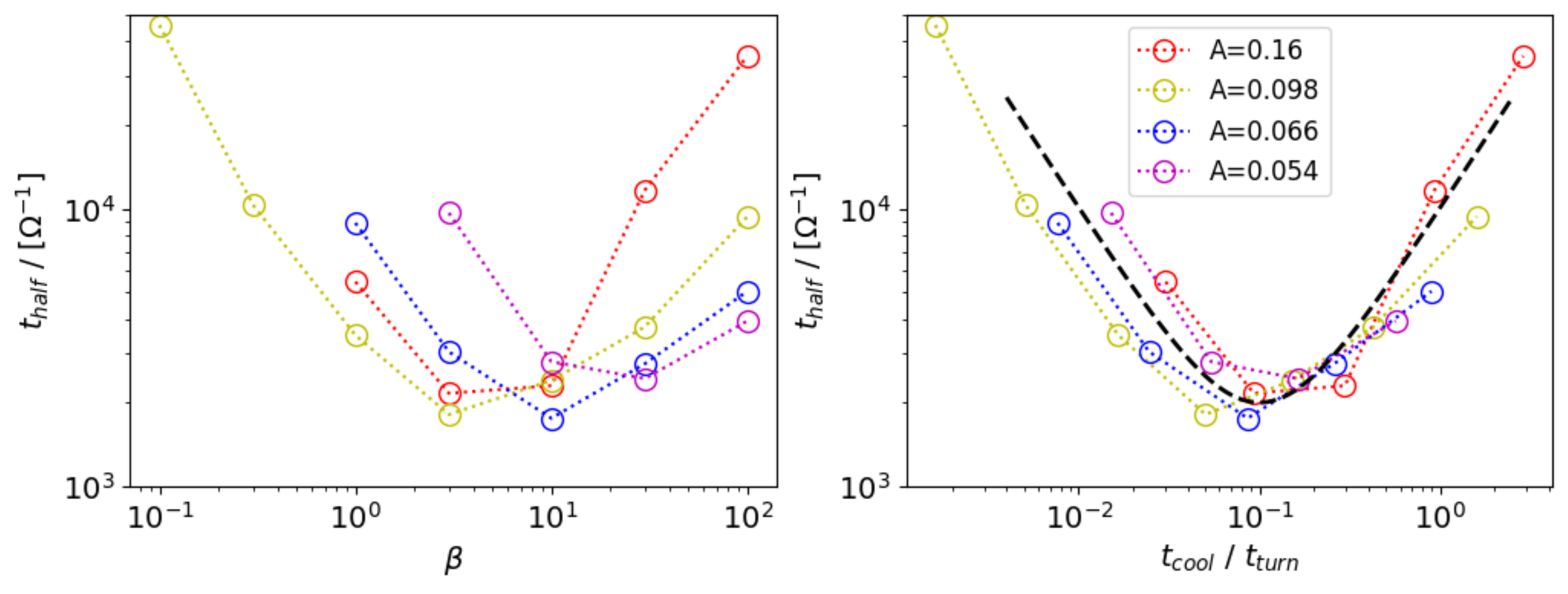}
    \caption{Half-life $t_{\rm half}$ of the vortices in our simulations. We show how it depends on $\beta$, the ratio of the cooling time to the dynamical time, for various values of $A$ on the left; and the same but replacing $\beta$ with $t_{\rm cool}/t_{\rm turn}$, the ratio of the cooling time to the vortex turnaround time, on the right. We find that $t_{\rm half}$ has a much stronger correlation with $t_{\rm cool}/t_{\rm turn}$, and can be approximated, in an order-of-magnitude sense, by the black dashed line in the right panel, described by Equation \ref{eq:fit}. When $t_{\rm cool}/t_{\rm turn}$ is near $0.1$, $t_{\rm half}$ is as short as $\sim2000\Omega^{-1}$, or about 300 orbits.}
    \label{fig:t_half}
\end{figure*}

To quantify the decay rate, we measure the half-life, $t_{\rm half}$, shown in their $\eta$ curves. By visual inspection, we locate a time, $t_0$, where a single vortex has formed and settled down in our simulation, then we measure $\eta$ at $t_0$, and find a subsequent time where $\eta$ falls to half its value at $t_0$. In cases where $\eta$ has not decreased by half by the end of our simulations, we extrapolate by estimating $t_{\rm half}$ as $(t_{\rm end}-t_0)/(\eta(t_{\rm end})-\eta(t_0)/2)$, where $t_{\rm end}=1000$ orbits is the simulation end time. The left panel of Figure \ref{fig:t_half} plots $t_{\rm half}$ as a function of $\beta$.

Whether cooling is fast or slow should be measured against the vortex turnaround time $t_{\rm turn}$ (see Section \ref{sec:metrics}). We measure $t_{\rm turn}$ at $t_0$, and show on the right panel of Figure \ref{fig:t_half} that $t_{\rm half}$ seems to follow a simple pattern with respect to $t_{\rm cool}/t_{\rm turn}$, independent of the initial amplitude of the vortex. The black dashed line is not a formal fit, but an order-of-magnitude approximation by eye that appears to match the empirical data quite well. It reads as follows:
\begin{equation}
    t_{\rm half} = 1000\,\Omega^{-1} \left[\frac{t_{\rm cool}}{0.1\,t_{\rm turn}} + \left(\frac{t_{\rm cool}}{0.1\,t_{\rm turn}}\right)^{-1}\right]\, .
    \label{eq:fit}
\end{equation}
The half-life of a cooling vortex can therefore be as short as $200\sim300$ orbits if $t_{\rm cool}$ is of order $0.1\,t_{\rm turn}$. Our measurements of $t_{\rm half}$ and $t_{\rm turn}$ are all summarized in Table \ref{tab:all}.

\begin{figure*}
    \centering
    \includegraphics[width=0.99\textwidth]{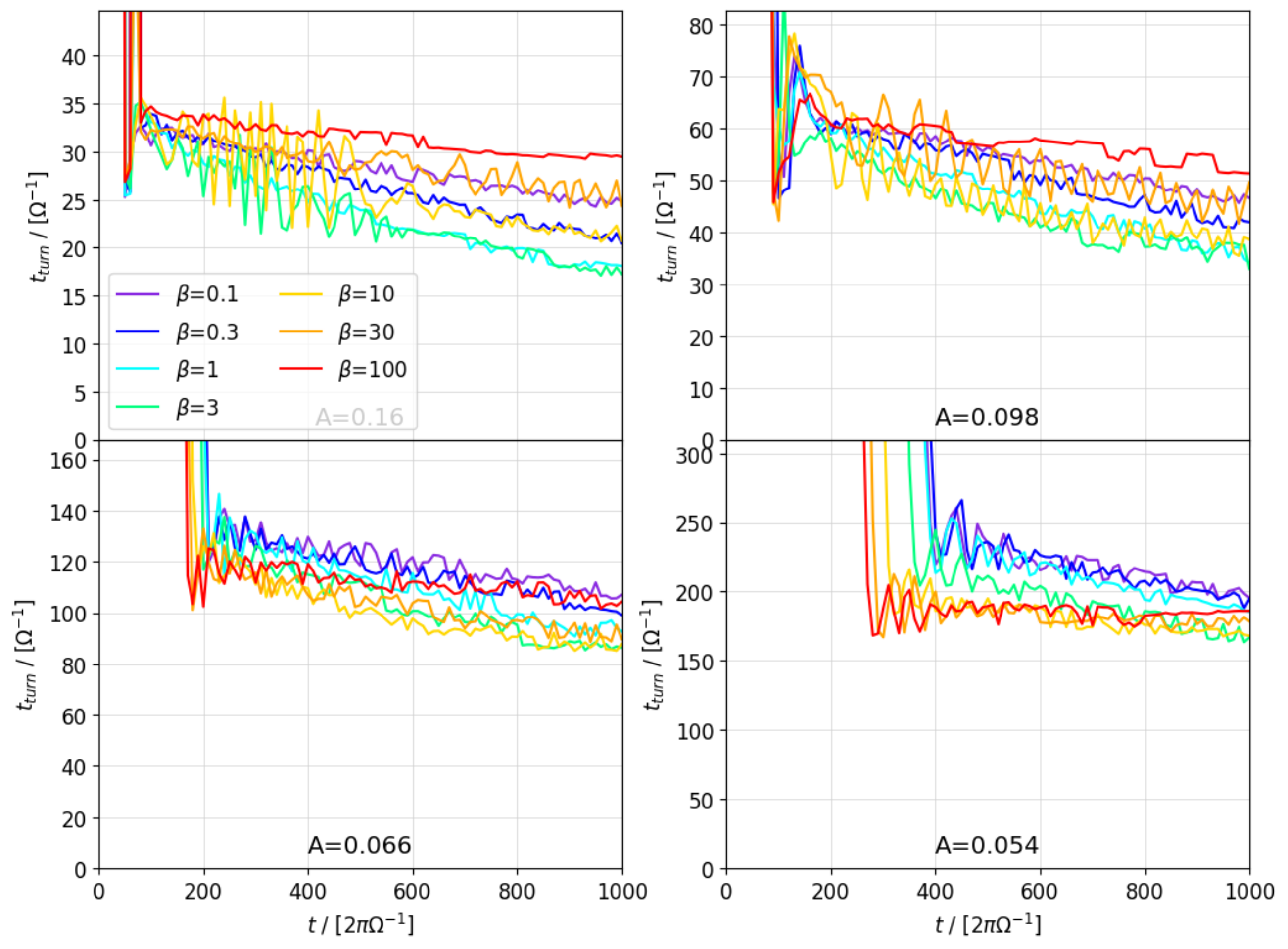}
    \caption{$t_{\rm turn}$ as a function of time for all 28 simulations, separated into 4 panels by the amplitude of the initial Gaussian peak $A$. Generally, $t_{\rm turn}$ decreases over time, in agreement with the vortex cores spinning up (e.g., Figure \ref{fig:2Dvor}). $t_{\rm turn}$ decreases more rapidly if $\beta$ is smaller. In fact, $t_{\rm turn}$ is nearly constant when $\beta=100$ for the weaker vortices.}
    \label{fig:t_turn}
\end{figure*}

One thing to note is that $t_{\rm turn}$ does evolve over time since the cores of our vortices are typically spinning up. Figure \ref{fig:t_turn} plots $t_{\rm turn}$ in all of our simulations, showing $t_{\rm turn}$ can decrease up to $\sim30\%$ over several hundred orbits, with a larger change when $\beta$ is smaller. This is consistent with the vortex evolution we observe in our simulations: as the vortices decay, their aspect ratios (how elongated they are) are also decreasing. Less elongated vortices have shorter $t_{\rm turn}$ (e.g., Appendix A of \citealt{Ono2018}). If Equation \ref{eq:fit} always applies, a constantly decreasing $t_{\rm turn}$ would imply an accelerating decay rate when $t_{\rm cool}<0.1\,t_{\rm turn}$, and decelerating decay when $t_{\rm cool}>0.1\,t_{\rm turn}$. Future, longer duration simulations may be able to confirm this.

Finally, we also note that there are some surprises and complications in our simulations. First, Figure \ref{eq:eta} shows that $\eta$ is not always decreasing---when $\beta\lesssim1$, we find that $\eta$ can in fact be increasing. It is not clear to us why, although it may simply have to do with imprecisions in our $\eta$ measurements as a result of the approximations employed (Section \ref{sec:metrics}). The vortices in these simulations do shrink slightly over our simulation time, but it is more than compensated by how much they spin up, leading to an overall increase in the Coriolis term $a v_{\rm rot}\Omega_{\rm K}$ in the definition of $\eta$ (Equation \ref{eq:eta}).

A perhaps related phenomenon is gap opening by the vortices. Vortices excite spiral waves at Lindblad resonances; this may weaken them but likely not substantially, since Figure \ref{fig:res} shows that in the absence of cooling, vortex strength is nearly constant over one thousand orbits. Although emitting these waves do not appear to affect vortex strength, when these waves damp, they deposit angular momentum locally, pushing gas away and opening a gap. These gaps alter the local vorticity profile and the effects of which are not well understood. Thermal relaxation affects this process by providing additional damping to the waves \citep{Miranda2020}, which are otherwise damped only by wave-steepening \citep{Rafikov02}.

\section{Conclusions and Discussions} \label{sec:conclude}

We have confirmed that anticyclonic vortices in Keplerian disks decay over time when the gas cooling time is finite. Using a set of 28 shearing box simulations, varying cooling time and initial condition, we have measured the decay time of vortices formed from RWI, and found that cooling-induced decay is most effective when $t_{\rm cool}$ has an intermediate value of $\sim0.1\,t_{\rm turn}$. At its fastest, the vortex half-life is as short as $200\sim300$ orbits. In our parameter space covering a range of different values of $t_{\rm turn}$, $t_{\rm cool}\sim0.1\,t_{\rm turn}$ corresponds to a $\beta$ value between 3 and 30.

Our results shed light on why annular structures appear to be more common than asymmetric structures in protoplanetary disks. Vortices may survive only thousands of orbits if cooling plays a role. The key question here is whether the cooling time in protoplanetary disks is within the optimal range for cooling-induced vortex decay. \cite{Zhu2015} estimated the disk cooling time using the gray atmosphere approximation of \cite{Hubeny1990}, and found that $\beta$ can range from $10^{5}$ to $10^{-2}$ between 1 to 100 au. For completeness, we restate their Equation 9 here:
\begin{equation}
    \beta \approx 0.07 \left(\frac{r}{\rm 30\,au}\right)^{-4.5} \left(\frac{T_{\rm disk}}{\rm 60\,K}\right)^{-3} \left(\frac{\kappa}{\rm 1\,cm^{2} \,g^{-1}}\right) \, ,
    \label{eq:tcool_thick}
\end{equation}
if the disk is optically thick, and
\begin{equation}
    \beta \approx 0.002 \left(\frac{r}{\rm 30\,au}\right)^{-1.5} \left(\frac{T_{\rm disk}}{\rm 60\,K}\right)^{-3} \left(\frac{\kappa}{\rm 1\,cm^{2} \,g^{-1}}\right)^{-1} \, ,
    \label{eq:tcool_thin}
\end{equation}
if it is optically thin. We have simplified their expression by assuming the disk surface density follows the minimum mass solar nebula \citep{Hayashi81}, and separating it into the optically thick and thin branches. In the equations, $\kappa$ is the Rosseland mean opacity and $T_{\rm disk}$ is either the disk midplane temperature or the equilibrium temperature set by stellar irradiation, whichever one is higher. By this estimate, cooling-induced decay may be effective around 10 au in optically thick disks. It is important to note, however, that this is an estimate of the cooling time in the disk midplane. For instance, passively heated, optically thick disks have shorter cooling times at higher altitudes. To complicate matters, global gradients also plays a role; for some disk profiles, thermal relaxation can trigger either the subcritical baroclinic instability \citep{Klahr2003,Klahr2004,Petersen2007a,Petersen2007b,Lesur2010,Lyra2011,Raettig2013,Barge2016}, which is a nonlinear instability that can amplify vortices, or the convective overstability \citep{Klahr14,Lyra14,Latter16}, which can generate vortices from linear perturbations. How these mechanisms may or may not operate in bumpy disks with narrows rings (potentially sandwiched by embedded planets) remains to be investigated.

Moreover, one should bear in mind that cooling is not the only cause of vortex decay; other avenues include the elliptical instability (see \citealt{Kerswell2002} for a review and references therein), dust feedback \citep{Johansen2004,Chang2010}, and 3D stratification effects \citep[][]{Barranco2005}. The elliptical instability is a robust 3D instability that leads to vortex decay under most circumstances \citep{Lesur2009}, although 3D baroclinicity can amplify the vortices and prolong their lifetimes \citep{Barge2016}. Dust grains are dragged into vortex cores and the back-reaction from the drag force can weaken, and eventually disperse, vortices. This mechanism is more effective if the system has a higher dust-to-gas ratio and if the dust grains are marginally coupled to the gas (i.e., Stokes number close to unity)\citep{Fu2014,Crnkovic2015}. Finally, vortices in 3D stratified disks may be ripped apart by vertical shear; and midplane vortices may even be unstable to the growth of an anti-symmetric mode that can destroy them \citep{Barranco2005}. The 2D, gas-only simulations in this paper are simplifications that do not capture any of the above effects.

\begin{figure*}
    \centering
    \includegraphics[width=0.99\textwidth]{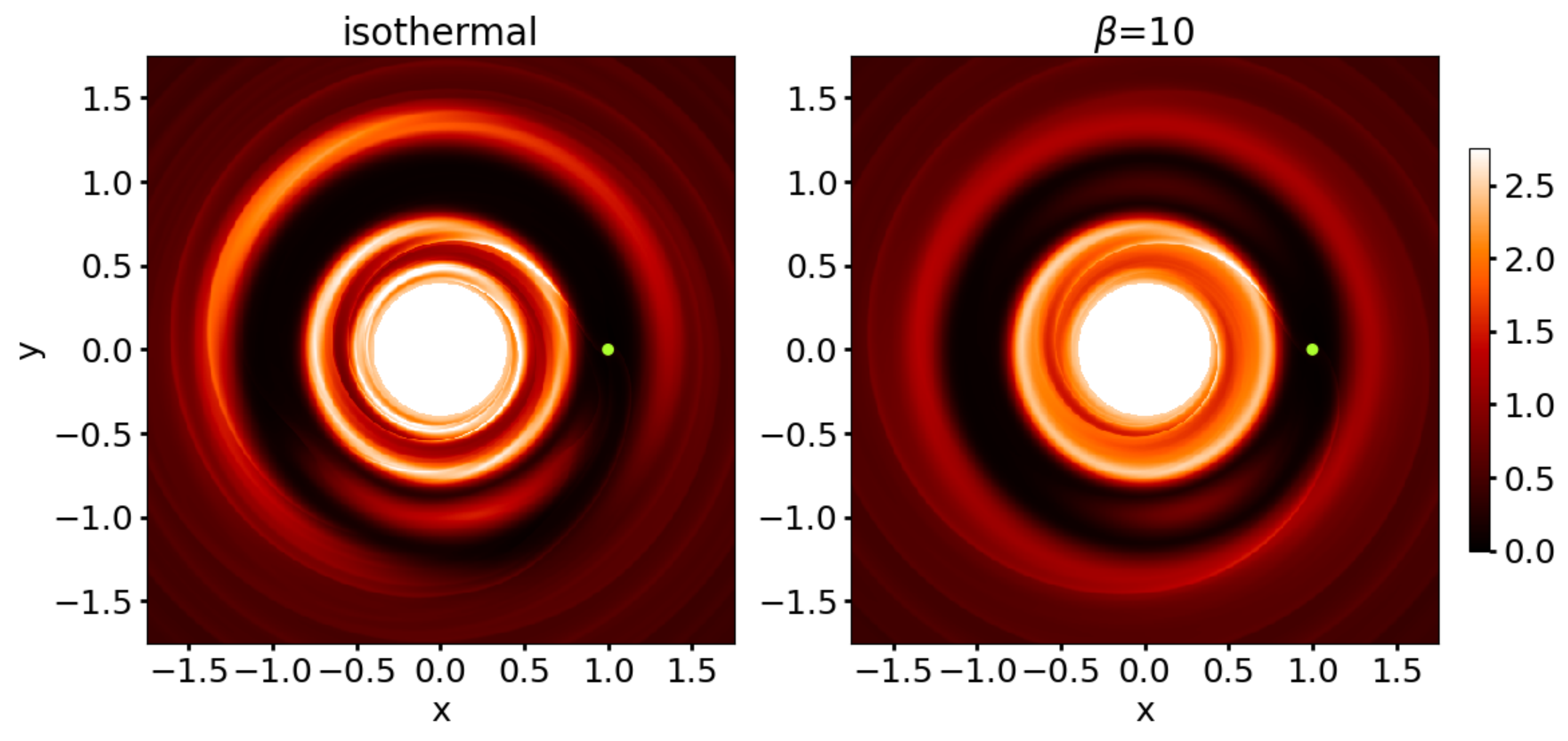}
    \caption{Surface density snapshots taken at 700 orbits after the planets are introduced. Saturn-mass planets are located at $r=1$, indicated by the green dots. Surface density is normalized such that in the absence of the planet, it equals 1 at $r=1$. Material evacuated from the planetary gaps pile up at gap edges to form rings. These rings are unstable to RWI, and turn into vortices. In the isothermal case (left), we see these vortices as clear azimuthal asymmetries. But with cooling (right), the vortices decay away and we recover axisymmetric rings.}
    \label{fig:planet}
\end{figure*}

Despite the simplicity of our model, cooling on its own may already be effective in explaining the rarity of large-scale asymmetries in disks. As a proof of concept, we additionally perform two global planet-disk interaction simulations to illustrate how cooling can restore stability in rings. The setup is nearly identical to that used by \cite{Fung2014}, but here the explicit viscosity in the disk is set to zero, and the resolution is increased to about 16 cells per scale height. A planet of Saturn's mass (planet-to-star mass ratio is $2.86\times10^{-4}$) is placed at $r=1$, where the disk's aspect ratio is $0.05$. We compare two cases: one isothermal and one with $\beta=10$. Figure \ref{fig:planet} plots surface density of these two models after 700 orbits. As with other simulations using low-viscosity or inviscid disks, vortices form at the edges of the planetary gaps \citep[e.g.,][]{deValBorro2007,Lin10,Yu10,Fung17}. When the gas is isothermal, these vortices persist and create large-scale asymmetries in the disks. But with cooling, we find that the vortices quickly decay away, and the end result is a system of rings reminiscent to those observed in protoplanetary disks.

Recently, \cite{vanderMarel2021} found that large-scale asymmetries in disks are more common in disks with lower gas densities. Their finding is in line with cooling-induced vortex decay if these asymmetric disks also have lower dust densities and are optically thick at the locations where the asymmetries are found. In optically thick disks, having less dust reduces the disk's optical depth and allows for more efficient radiative cooling. In turn, faster cooling leads to slower decay if $t_{\rm cool}<0.1\,t_{\rm turn}$. On the scale of tens of au, $t_{\rm cool}$ is likely shorter than the dynamical time $\Omega^{-1}$ (i.e., Equation \ref{eq:tcool_thick}). Meanwhile, the turnaround time of a vortex is typically tens of $\Omega^{-1}$ (Table \ref{tab:all}). This places us firmly in the $t_{\rm cool}<0.1\,t_{\rm turn}$ regime. Vortices may form in most disks at tens of au, but they decay away more quickly in disks with higher dust (and gas) densities.

\acknowledgments
We thank Eugene Chiang and Jim Stone for helpful discussions. An anonymous referee provided insightful comments that significantly improved this manuscript. JF gratefully acknowledges support from the Institute for Advanced Study. TO acknowledges the use of the \texttt{Athena++} code \cite{Stone2020} and Cray XC50 at Center for Computational Astrophysics, National Astronomical Observatory of Japan for some of the analysis in this work. This work was partially supported by Japan Society for the Promotion of Science (JSPS) KAKENHI Grant Numbers 20J01376 (TO).

\bibliography{Lit}{}
\bibliographystyle{aasjournal}

\end{document}